# Broadband light extraction from near-surface NV centers using crystalline-silicon antennas

*Minjeong Kim[1], Maryam Zahedian[1], Wenxin Wu[1], Chengyu Fang[1], Zhaoning Yu[2], Raymond A. Wambold[1], Ricardo Vidrio[1], Yuhan Tong[1], Shenwei Yin[1], David A. Czaplewski[4], Jennifer T. Choy[1,2,3]\*, Mikhail A. Kats[1,2]\**

\*mkats@wisc.edu, jennifer.choy@wisc.edu

[1] Department of Electrical and Computer Engineering, University of Wisconsin–Madison, Madison, WI 53706, USA

[2] Department of Physics, University of Wisconsin–Madison, Madison, WI 53706, USA

[3] Center for Nanoscale Materials, Argonne National Laboratory, Lemont, IL 60439, USA

**ABSTRACT**

We use crystalline silicon (Si) antennas to efficiently extract broadband single-photon fluorescence from shallow nitrogen-vacancy (NV) centers in diamond into free space. Our design features relatively easy-to-pattern high-index Si resonators on the diamond surface to boost photon extraction by overcoming total internal reflection and Fresnel reflection at the diamond-air interface, and providing modest Purcell enhancement, without etching or otherwise damaging the diamond surface. In simulations, ~17 times more single photons are collected from a single NV center compared to the case without the antenna; in experiments, we observe an enhancement of ~9 times, limited by spatial alignment between the NV and the antenna. Our approach can be readily applied to other color centers in diamond, and more generally to the extraction of light from quantum emitters in wide-bandgap materials.

Color centers in materials such as diamond,[1–3] silicon carbide,[4,5] and hexagonal boron nitride (h-BN)[6,7] are components of a host of emerging quantum technologies, including quantum networking,[8–10] computation,[11–13] and sensing of electric and magnetic fields,[14,15] temperature,[16,17] strain,[18,19] and inertial motion.[20,21] Because these materials tend to have relatively high refractive index, light extraction from color centers is an important engineering problem; for example, only ~3% of the light emitted from negatively-charged nitrogen-vacancy (NV) centers (Fig. 1(a)) in (100) bulk diamond (refractive index n ~ 2.4) is transmitted through the diamond-air interface due to a combination of total internal reflection (TIR) and Fresnel reflection at the diamond-air interface (Fig. 1(b)).

Several solutions to the light-extraction problem have been explored, involving etching of the diamond surface, such as creating parabolic reflectors,[22] nanowires followed by metallization,[23] gratings,[24,25] and metasurfaces.[26] However, etching of the diamond surface can significantly affect the quality of NV centers just below the surface through material damage to the crystal or introduction of surface defects, but these near-surface NV centers are otherwise ideally suited for certain sensing applications such as high-resolution measurement of fields[27–29] and spin-based chemical sensing.[30] Furthermore, while etching of diamond is now reasonably well established,[31–

[33] it remains technically difficult and often requires a dedicated etching chamber. Therefore, the ideal solution for light extraction from NV centers in diamond and other color centers, especially for sensing applications, is to build structures on top of the diamond without etching or otherwise modifying the diamond surface. Two proposals for such light extractors have used adjoint optimization to design sophisticated high-index structures with small features and high aspect ratios that can be fabricated on top of a diamond surface: one experimentally realized structure in gallium phosphide,[34] and one proposal in silicon.[35] Here, we demonstrate relatively easy-to-design and single-step fabricated crystalline-silicon antennas that can efficiently extract and beam broadband fluorescence from NV centers, substantially increasing the number of single photons that can be collected with an objective lens in free space.

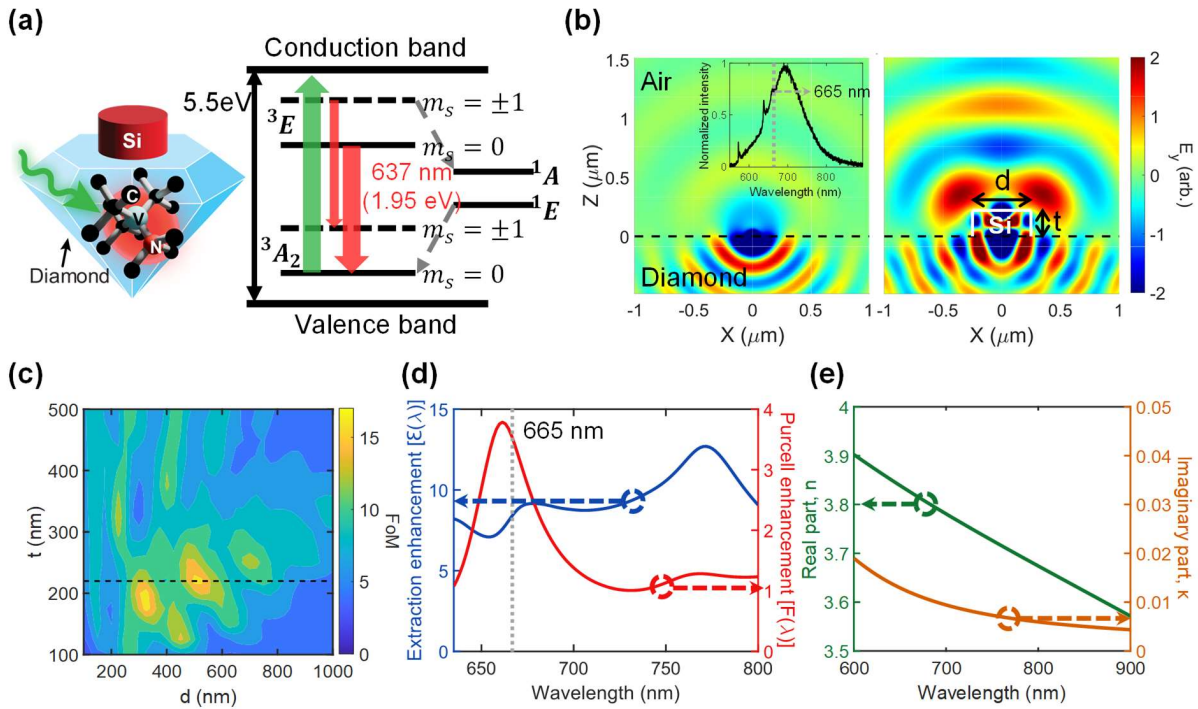

**Figure 1. (a)** Visualization of the NV⁻ center in diamond and its energy-level diagram. The green arrow indicates the excitation, and the red arrows indicate the radiative transitions. **(b)** Snapshot of the simulated electric fields near a y-polarized dipole (parallel to interface) 10 nm below the interface between diamond and air, taken along a slice in the XZ plane at y = 0 (the center of the device and the y-coordinate of the NV) at a wavelength of 665 nm. The Si pillar is outlined. d: diameter of the pillar; t: thickness of the pillar. (Inset) Normalized photoluminescence spectrum of the NV center in bare diamond.

**(c)** Simulated color map of the figure of merit (FoM) with respect to the diameter ($d$) and thickness ($t$) of the Si pillar ($d$, $t$ resolution of 25 nm). The dotted line indicates the thickness of the Si device layer in this work (220 nm). **(d)** Simulated extraction enhancement [$\mathcal{E}(\lambda)$] and Purcell enhancement [$F(\lambda)$] of the Si antenna for NV depth of 10 nm below the diamond/air interface (blue line) with Si pillar with d = 500 nm and t = 220 nm. Purcell enhancement (red line) of the Si pillar. Both extraction enhancement and Purcell enhancement are averaged across two dipole orientations, [$\overline{11}2$] and [$1\overline{1}0$]. **(e)** Real and imaginary parts of the refractive index of crystalline Si, measured using variable-angle ellipsometry.

Fig. 1(a) shows our design, featuring a cylindrical pillar of single-crystal silicon (Si) positioned on the diamond surface, immediately above a sub-surface NV center at a depth of 5–15 nm. We utilized single-crystalline Si because its indirect bandgap results in sufficiently low absorption losses at wavelengths longer than ~600 nm, which correspond to the NV center spectrum, including the zero-phonon line and the phonon sideband. In contrast, polycrystalline and amorphous Si exhibit higher absorption coefficients at these wavelengths.[36] We validated our design with finite-difference time-domain (FDTD) simulations (Ansys Lumerical FDTD). In the simulations (Fig. 1(b)), we positioned a y-polarized electric dipole 10 nm below the diamond-air interface (note that the dipole orientations with respect to the interface depend on the NV orientation and on the diamond cut),[37] observing that the majority of the emitted light remains confined within the diamond. However, incorporating a single-crystal Si pillar allows us to overcome total internal reflection (TIR) and Fresnel reflection due to the overlap of the near field from the emitter with the high-index Si resonator, thereby significantly enhancing light extraction from the dipole in the diamond (Fig. 1(b)). To optimize the diameter ($d$) and thickness ($t$) of the Si pillar for maximum light extraction across the NV emission spectrum, we use a spectrum-averaged figure of merit (FoM), which quantifies how much more light emerges from the diamond [35]:

$$FoM = \frac{\int I_{NV}(\lambda) \cdot \mathcal{E}(\lambda) \cdot F(\lambda) d\lambda}{\int I_{NV}(\lambda) d\lambda} \qquad (1)$$

Here, $I_{NV}(\lambda)$ is the normalized emission spectrum of NV centers in bulk diamond,[38] and $\mathcal{E}(\lambda)$ is the extraction enhancement of our device, defined as the ratio of the number of photons emitted into free space in the presence of the Si pillar to the number of photons emitted from the bare diamond surface without the Si pillar, accounting only for collection enhancement and not the Purcell enhancement. The bounds of the integral in Eq. (1) are set to 635–800 nm to encompass the NV zero-phonon line and phonon-sideband emission.

We swept over the Si pillar thickness ($t$) and diameter ($d$), finding several local maxima (Fig. 1(c)) and, in particular, a local maximum for $d \sim 500$ nm and $t \sim 220$ nm, yielding a FoM of approximately 17, indicating an increase of the optical power emitted into the air by a factor of 17 compared to the bare diamond surface. 220 nm is also a common silicon device-layer thickness used in silicon photonics,[39,40] and therefore crystalline Si membranes of this thickness are readily commercially available. The extraction enhancement $\mathcal{E}(\lambda)$ and Purcell enhancement $F(\lambda)$ for the optimal Si pillar are shown in Fig. 1(d), calculated by averaging the emitted power over two orthogonal electric dipoles at the NV location.[37] The circular shape of the pillar ensures robustness to all four possible orientations of the NV in (100)-oriented diamond. The Purcell enhancement, representing the increased spontaneous emission rate experienced by the dipole,[41] exceeds 2 for wavelengths in the 645–685 nm range.

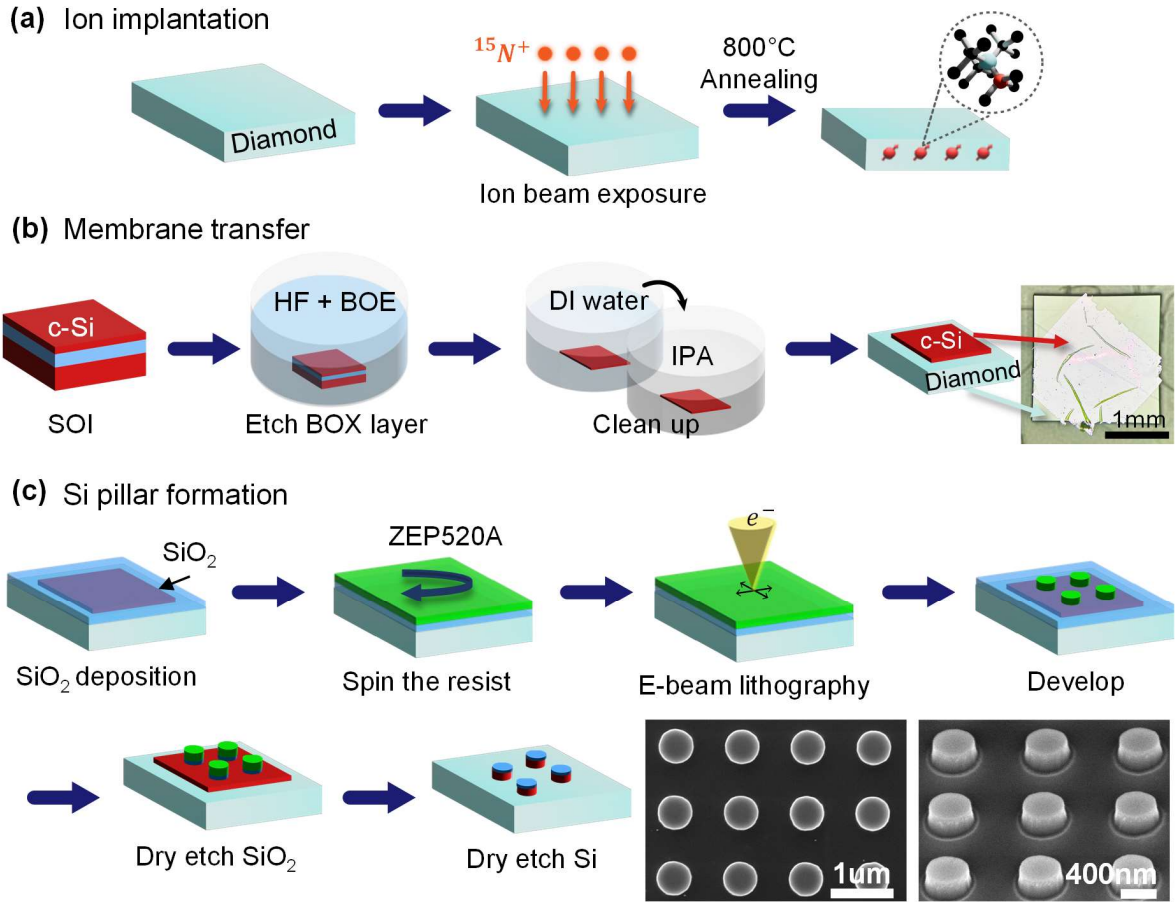

**Figure 2**. The fabrication procedure, along with photo and SEM images of the fabricated Si antennas. **(a)** The diamond sample is implanted with nitrogen ions at 4 keV and a dose of $2 \times 10^9$ N/cm$^2$, and annealed to generate NV centers roughly 7 nm below the surface (Supporting Fig. S1). **(b)** The buried oxide (BOX) layer of the SOI piece is undercut using a hydrogen fluoride (HF) and buffered oxide etchant (BOE) solution, and the Si membrane is wet-transferred onto the diamond substrate. The inset on the left is an optical microscope image of a c-Si membrane transferred to a 2 mm x 2 mm diamond sample. **(c)** An SiO$_2$ hard mask is deposited using plasma-enhanced chemical vapor deposition (PECVD). ZEP520A resist is spun onto the substrate, patterned via e-beam exposure, and developed. The SiO$_2$ and Si layers are then etched to form Si pillars, as shown in the SEM images.

We fabricated the Si pillars by transferring a single-crystal silicon membrane from a silicon-on-insulator (SOI) wafer onto the diamond surface using an epitaxial lift-off technique.[42,43] A chemical vapor deposition (CVD) diamond sample was ion-implanted with nitrogen ions, followed by vacuum annealing at 800 °C and tri-acid cleaning (Fig. 2(a)). The Si membrane was then released and transferred onto the diamond (Fig. 2(b)), and the structure was then defined using electron-beam lithography, etching of the hard mask, and Si etching (Fig. 2(c)). To protect the diamond surface from O₂ plasma etching[31] and minimize unwanted contamination, particularly for near-surface NV centers (<10 nm), we intentionally left a thin layer of Si on top of the diamond during the etching process, targeting a thickness of 20–30 nm (Supporting Fig. S12). The SEM images of the fabricated Si antennas are shown in Fig. 2(c). Detailed fabrication procedures are provided in the supporting information.

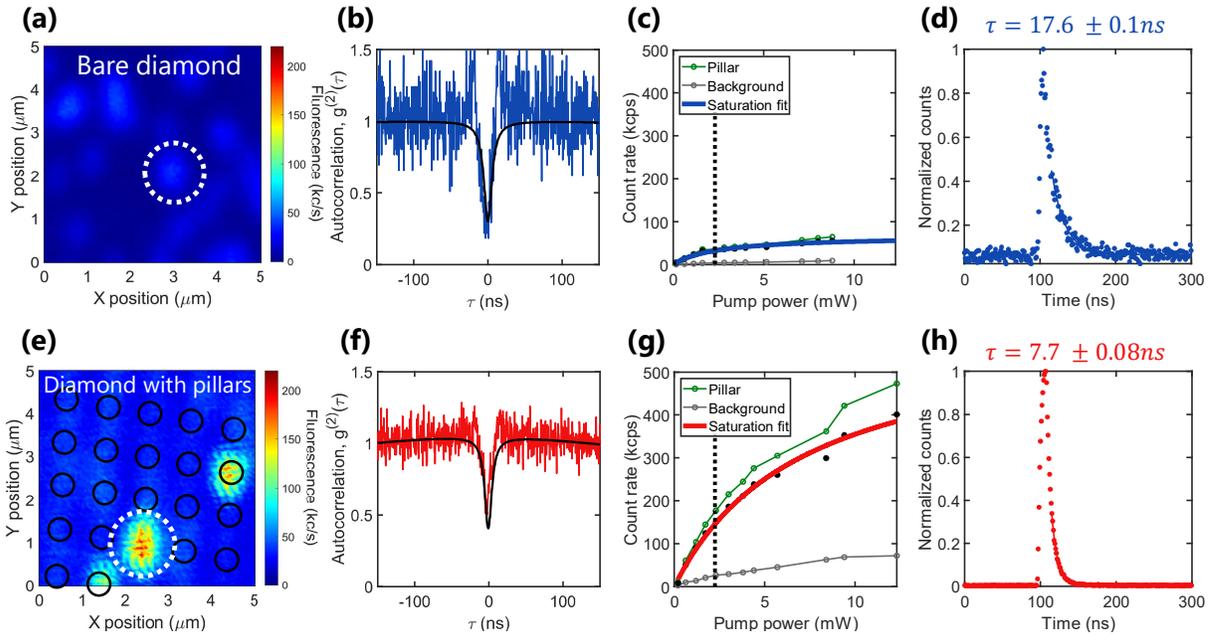

**Figure 3.** Comparison of NVs from bare diamond (a-d) and with Si antennas (e-h). The NV depth is similar between the two samples: ~9 nm in a-d, and ~7 nm in e-h. **(a, e)** Confocal microscopy scans of a 5-μm by 5-μm region, with the circled regions (white dashed circles) representing the spots used for measurement and the black circles representing other nearby pillars. **(b, f)** Autocorrelation function $g^2(\tau)$ for a single NV in the bare diamond (b), and underneath a Si pillar (f), showing anti-bunching at zero time delay $\tau$, indicating emission of non-classical light. No background correction is performed. **(c, g)** Saturation curves showing the total count rates as a function of pump power. Subtraction of the background from the total yields the isolated NV emission (black dots), which are fitted to a saturation model (blue and red lines, respectively). The vertical black dotted line denotes the pump power used during the confocal scan, autocorrelation, and lifetime measurements. **(d, h)** Normalized fluorescence decay for a single NV center in bare diamond, vs. with a Si antenna, measured at the specific points marked by the white dashed circles in (a) and (e). The decay curves were fitted to a one-exponential model. The fits yield time constants of 17.59 ± 0.09 ns for single NV in bare diamond (without Si antennas), and 7.7 ± 0.08 ns for single NV underneath a Si antenna.

The fluorescence measurements were performed by exciting the sample with a continuous wave laser at 515 nm and collecting the NV center emission through the same objective (numerical aperture NA = 0.95), using a home-built confocal microscope. The fluorescence was focused onto a single-mode fiber connected to an avalanche photodiode (APD). Confocal scan images of each sample from the APD output are displayed in Fig. 3(a) and (e). Note that the ion implantation energy for the bare diamond was 6 keV, while for the diamond with Si antennas, it was 4 keV, resulting in slightly different NV depths of ~9 nm for the bare diamond and ~7 nm for the diamond with Si pillars, but this difference does not significantly affect the conclusion (Supporting Fig. S1). As shown in Fig. 3(a) and (e), the emission from the NV center with a Si antenna is clearly enhanced compared to the bare diamond sample. To confirm single-photon characteristics of individual NV centers from both diamond samples, we measured the second-order autocorrelation function[44], $g^{(2)}(\tau) < 0.5$, without doing background subtraction, demonstrating that each contained a single-photon emitter without too much background (Fig. 3(b) and (f)). For both samples, we observed the characteristic optically detected magnetic resonance (ODMR) dip at 2.87 GHz[45] without an external magnetic field (Supporting Fig. S2).

Fluorescence intensity saturation curves of both samples were measured by varying the pump power (Fig. 3(c) and (g)). For the bare diamond, the background was measured in regions without NV centers, and for the diamond with Si antennas, the background was measured at a Si pillar without NV centers. The background-subtracted count rate, representing the NV fluorescence count rate, was fitted to the following saturation model[44]:

$$C(P) = \frac{C_{sat}}{1 + \frac{P_{sat}}{P}} \quad (2)$$

Here, $C_{sat}$ is the saturated count rate, $P_{sat}$ is the saturation power, and $P$ is the laser power. We measured the $C_{sat}$ of the NV center in the bare diamond to be ~67 kcps, while that for the NV center with the Si antenna is ~609 kcps, indicating a 9.1-fold increase. Meanwhile, $P_{sat}$ values in two cases differ: $P_{sat}$ of the NV center in the bare diamond is ~2.4 mW, and $P_{sat}$ of the NV center with the Si antenna is ~7.16 mW, though the comparison is not informative due to the complex spatial distribution under the Si pillar of the enhanced incident light (Supporting Fig. S7). The normalized fluorescence count time traces were fitted to a one-exponential decay model (see *Methods*), yielding a time constant of $17.59 \pm 0.09$ ns for the bare diamond. NVs near the diamond-air interface can exhibit longer lifetimes (~18 ns) compared to those in bulk diamond (~12 ns) due to the reduced emitted power caused by interference effects near the diamond-air interface [37]. The lifetime for the diamond with Si antennas is $7.7 \pm 0.08$ ns, indicating a 2.3-fold reduction in the lifetime in the presence of the Si antenna.

The 9.1-fold increase in measured broadband single-photon fluorescence is a significant improvement, but is below the FoM ~ 17 observed in simulations. We believe that the discrepancy is due to two factors: a potential difference in NV depth (Supporting Fig. S1)[46] and misalignment of the Si pillar with the NV center. The misalignment is to be expected given the density of NVs in our sample, the density of the pillar array we fabricated, and the lack of intentional alignment prior to the lithography (Supporting Fig. S5). We also note that the tilted sidewalls observed in the SEM image (Supporting Fig. S14), are expected to slightly reduce the FoM from ~17 to ~16. Due to the lack of precise alignment, it is also difficult to determine whether the incident field is enhanced or suppressed, making it challenging to accurately assess the lateral displacement between the Si pillar and the NV center (Supporting Fig. S6, S7).

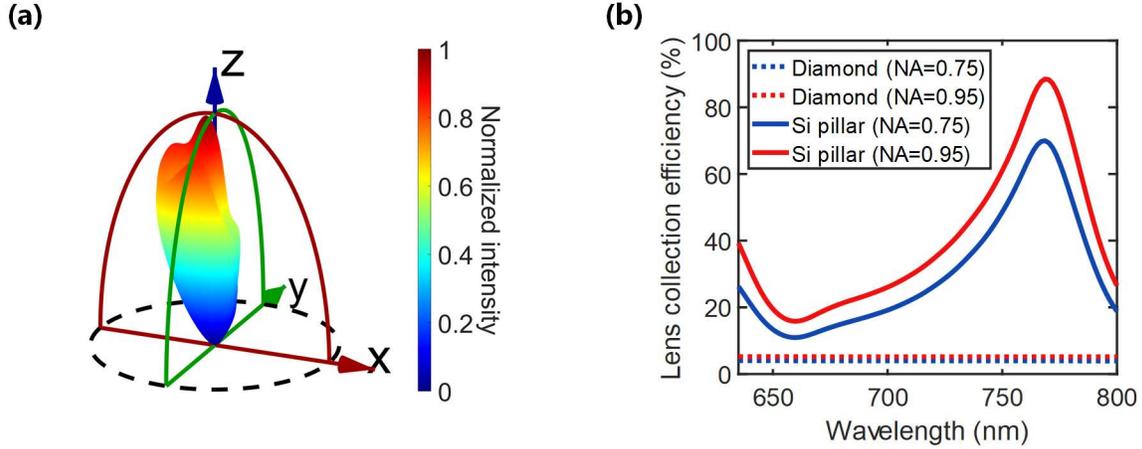

**Figure 4. (a)** Simulated far-field $|E|^2$ intensity distribution of the light emitted from NV center, averaged over the two dipole orientations $[\bar{1}\bar{1}2]$ and $[1\bar{1}0]$, at a depth of 10 nm passing through Si antenna. The plot is shown at a wavelength of 665 nm. **(b)** Calculated collection efficiency of the emitted light by a lens, defined as the fraction of emitted power from Si antenna collected by a lens with a specific numerical aperture (NA**)**. The increase in lens collection efficiency compared to bare diamond is due to the collection enhancement and does not include the Purcell enhancement**.** The lens used in the confocal microscope has an NA of 0.95.

In our previous work, where we generated a complex adjoint-optimized structure for light extraction,[35] the target performance was beaming into a specific Gaussian free-space mode of a certain size and orientation, which necessitated optimization over many structural degrees of freedom. However, the resulting structure was challenging to fabricate, and also was not rotationally symmetric, which meant that it was optimized for one particular NV orientation. Here, instead, we demonstrated that a much-simplified and rotationally symmetric design can achieve similar levels of light extraction, while sacrificing some beam quality (Fig. 4(a)). Despite the lower beam quality, optical collection can remain high—for example, for a free-space objective with NA of 0.95, as used in our confocal microscope setup, more than 40% of the broadband optical power emitted by the NV can be collected in free space.

In summary, we designed, fabricated, and characterized crystalline silicon (Si) antennas to enhance light extraction from nitrogen-vacancy (NV) centers in diamond. By leveraging the high refractive index of Si, our design achieves a simulated 17-fold enhancement compared to bare bulk diamond, while experimental measurements showed a 9.1-fold enhancement, with the difference due to experimental misalignment between the NV and the antenna. Our approach avoids the etching of the diamond surface, which can increase surface roughness or alter chemical termination,[47–49] preserving the integrity of near-surface NV centers, and making it well-suited for integrating nanophotonic structures with diamond. The Si platform can be further optimized and applied to other color centers in diamond, such as silicon-vacancy[50,51] and germanium-vacancy centers,[52,53] or to other quantum emitters in wide-bandgap materials.

**METHODS**

FDTD simulations

We performed 3D finite-difference time-domain (FDTD) simulations using Lumerical FDTD. The simulated structure consists of a Si pillar on polished diamond. The simulations incorporated one of two orthogonal dipoles, representing NV center emission, positioned 10 nm below the diamond-air interface, with wavelengths ranging from 635 nm to 800 nm. Both extraction enhancement and Purcell enhancement were calculated as the average of the two dipole orientations, $[\bar{1}\bar{1}2]$ and $[1\bar{1}0]$. The simulation region size was set to 3 μm × 3 μm × 4 μm. Mesh overrides were applied to the Si pillar with a uniform mesh size of 5 nm in the x, y, and z directions, while a non-uniform mesh was used elsewhere. For example, near the diamond-air interface, the mesh size was maintained at 5 nm, increasing gradually to 35 nm in the z-direction at heights exceeding 1.5 μm above the interface. In the x and y directions, the mesh size increased from 5 nm at the boundaries of the Si pillar to 20 nm at the edges of the simulation domain. The simulation boundaries were defined using perfectly matched layers (PML). Convergence testing was conducted to validate the accuracy and reliability of the simulation parameters.

Diamond preparation

The 2 mm x 2 mm chemical vapor deposition (CVD) electronic-grade (100) single crystalline diamond substrates with less than 0.03 ppm nitrogen concentration were used to create NV centers. For the bare diamond, the sample was ion-implanted with nitrogen ions at an energy of 6 keV, followed by vacuum annealing at 800 °C and tri-acid cleaning with equal parts nitric ($HNO_3$), perchloric ($HClO_4$), and sulfuric acids ($H_2SO_4$). For the diamond with Si antennas, a similar substrate underwent strain relief etching using inductively coupled plasma reactive-ion etching (ICP-RIE, Plasma-Therm),[47] ion implantation at an energy of 4 keV, followed by vacuum annealing at 800 °C and tri-acid cleaning.

Device fabrication

To confirm the optical properties of the crystalline Si from SOI wafers, we measured real and imaginary parts of refractive index of the device layer from an SOI wafer using variable-angle spectroscopic ellipsometry (J.A. Woollam V-VASE) (Fig. 1(e)). Then, SOI wafers with a 220 nm layer of single-crystal silicon were diced, cleaned, and the silicon membranes were released using a mixture of hydrogen fluoride (HF) and buffered oxide etchant (BOE). The membranes were then transferred onto the diamond substrate (Fig. 2(b)). A ~27 nm layer of $SiO_2$ was deposited onto the silicon membrane using PECVD as a hard mask. Arrays of circular patterns with a diameter of 500 nm were defined using electron-beam lithography. These patterns were transferred into the hard mask using ICP-RIE (Oxford) with $CHF_3$ and $O_2$. Finally, 220 nm tall silicon antennas were created using HBr and $O_2$ etching (Fig. 2(c)). The $SiO_2$ hard mask was not removed, and is not expected to significantly affect the optical properties due to its low index and small thickness (Supporting Fig. S12).

Measurement

Confocal scans, ODMR, and lifetime measurements were performed using a custom-built confocal fluorescence microscope. NV centers were excited with a 515 nm laser (Omicron LuxX 515-100), and emission was collected using two APDs (Laser Components COUNT-100N-FC) with filters through a 0.95 NA air objective at ~2.3 mW. A fast-steering mirror (FSM-300-NM) was used for scanning, Continuous-wave 515 nm laser was used for confocal scans and ODMR, with fluorescence images captured using Qudi software, scanning 5 μm x 5 μm regions at 0.1 μm resolution. For ODMR, microwaves were applied via a printed circuit board (PCB), sweeping from 2.83 GHz to 2.92 GHz. Lifetime measurements were conducted using ~5 ns pulse excitations generated via digital control of the laser using a Pulse Streamer (Swabian instruments), with time-gated measurements recorded by a Time Tagger (Swabian instruments) at ~2.3 mW (in CW mode). Contributions from the instrument response and background fluorescence to the time-binned signal are constrained to the first ~10 ns after excitation, whereas the NV center emissions from bare diamond or Si antennas are associated with the longer-lived component. To determine the NV lifetimes, we start fitting the time-resolved fluorescence traces after ~10 ns to a single-exponential model.

## AUTHOR INFORMATION

**Corresponding Author**

*Mikhail A. Kats - Department of Electrical and Computer Engineering, University of Wisconsin–Madison, Madison, WI 53706, USA; Department of Physics, University of Wisconsin–Madison, Madison, WI 53706, USA; E-mail: mkats@wisc.edu

*Jennifer T. Choy - Department of Electrical and Computer Engineering, University of Wisconsin–Madison, Madison, WI 53706, USA; Department of Physics, University of Wisconsin–Madison, Madison, WI 53706, USA; E-mail: jennifer.choy@wisc.edu

**Funding Sources**

Contributions from MJK, ZY, and MAK were supported primarily by the National Science Foundation under Grant No. CHE-1839174, and in the later stages by the U.S. Department of Energy Office of Science National Quantum Information Science Research Centers (Q-NEXT). Contributions from MZ and JTC are supported by the U.S. Department of Energy, Office of Science, Basic Energy Sciences under Award #DE-SC0020313. Contribution from WW is supported by the University of Wisconsin – Madison Office of the Vice Chancellor for Research and Graduate Education with funding from the Wisconsin Alumni Research Foundation.

**ACKNOWLEDGMENT**

The authors thank Robert Hamers discussions early on in the project. The majority of the fabrication work was performed at the Center for Nanoscale Materials, a U.S. Department of Energy Office of Science User Facility, was supported by the U.S. DOE, Office of Basic Energy Sciences, under Contract No. DE-AC02-06CH11357. The authors also acknowledge use of facilities and instrumentation at the UW-Madison Wisconsin Centers for Nanoscale Technology (wcnt.wisc.edu) partially supported by the NSF through the University of Wisconsin Materials Research Science and Engineering Center (DMR-2309000).

**Ion implantation and diamond preparation**

*Single NV centers in the bare (100) single crystalline diamond:* A 2 mm × 2 mm piece of electronic-grade diamond with a nitrogen concentration of less than 0.03 ppm was used as the reference sample. The diamond underwent tri-acid cleaning with a solution consisting of equal parts nitric, perchloric, and sulfuric acids, and was then implanted with a fluence of $2 \times 10^9$ N/cm² and an implantation energy of 6 keV. The implanted diamond was then annealed under high vacuum (<$10^6$ Torr) at 800 °C to form NV centers centered a depth of ~9.4 nm (Supporting Fig. S1), followed by another tri-acid cleaning.

*Single NV centers in (100) single crystalline diamond with Si antennas:* A 2 mm x 2 mm piece of electronic-grade diamond with a nitrogen concentration of less than 0.03 ppm (ELSC20, Thorlabs) was used for implantation. The sample underwent strain-relief etching using inductively coupled plasma reactive ion etching (ICP-RIE, Plasma-Therm), resulting in the removal of ~5 µm from the surface, followed by tri-acid cleaning. Ion-beam exposure with a fluence of $2 \times 10^9$ N/cm² and an implantation energy of 4 keV was performed. The implanted diamond was annealed under high vacuum (<$10^6$ Torr) at 800 °C to create NV centers centered at ~7 nm (Supporting Fig. S1), based on SRIM simulation. Post annealing, tri-acid cleaning was conducted.

**Device fabrication**

*Preparation of silicon membrane and transfer to diamond:* SOI wafers, with a 220 nm layer of single-crystal silicon, were diced into 1 mm x 1 mm squares. These pieces were cleaned in Remover 1165 at 100 °C for 5 minutes, followed by ultrasonication in acetone and isopropanol (IPA) for 5 minutes each. The SOI piece was then immersed in a 1:1 mixture of 49% hydrogen fluoride (HF) and 20:1 buffered oxide etchant (BOE) for 17 hours to release the silicon (Si) membrane. The floating Si membrane was transferred to a 6:1 BOE solution for 4 hours to remove any remaining $SiO_2$ residue, followed by thorough rinsing in deionized (DI) water and IPA to clean up any HF residue. The ion-implanted diamond was cleaned in acetone and IPA for 5 minutes each. The silicon membrane was scooped up with a diamond and dried with a nitrogen gun. The diamond with the transferred silicon membrane was then placed on a hot plate at 90 °C for 10 minutes, followed by rapid thermal annealing (RTA) at 350 °C in nitrogen for 5 minutes.

*Patterning and fabrication of Si antennas:* The diamond with the transferred Si membrane was cleaned in a piranha solution, 4:1 mixture of sulfuric acid ($H_2SO_4$) and hydrogen peroxide ($H_2O_2$) for 2 minutes, followed by ultrasonication in acetone and IPA for 2 minutes each. A ~27 nm layer of $SiO_2$ was deposited onto the silicon membrane using PECVD (Oxford Plasmalab 100). A 1:1 mixture of ZEP520A and anisole was spin-coated onto the sample and baked at 150 °C for 3 minutes. Arrays of circular patterns were then defined using a JEOL JBX-8100FS electron-beam lithography system at 100 keV. After exposure, the pattern was developed in n-Amyl acetate for 1 minute and then rinsed with IPA. The features were defined in the $SiO_2$ hard mask using inductively coupled plasma reactive-ion etching (ICP-RIE, Oxford) with $CHF_3$ and $O_2$. Finally, 220-nm-tall silicon antennas were created using HBr and $O_2$ etching, and any remaining resist residue was removed with an $O_2$ plasma treatment.

**Experimental details of the optical measurements**

Confocal scans, optically detected magnetic resonance (ODMR), and lifetime measurements for diamond samples were performed using a home-built confocal fluorescence microscope. NV centers were excited by a 515 nm green laser (Omicron LuxX 515-100), and the emitted fluorescence was collected using two avalanche photodiodes (APDs, Laser Components COUNT-100N-FC), after passing through a 552 nm and 660 nm edge dichroic beam splitter and a 650 nm long-pass filter. This setup ensures that any Raman peaks arising from the interaction between the 532 nm and diamond (the most prominent peaks occurring around ~575 nm and ~620 nm), along with contributions from the neutral NV center, are filtered out to avoid noise in the fluorescence signal. Both excitation and emission were focused and collected by an air objective with a numerical aperture (NA) of 0.95 (CFI Plan Apo Lambda 40XC/0.95 Air). A fast-steering mirror (FSM-300-NM) was employed to scan the laser beam across the sample. For lifetime measurements, excitation with ~5 ns pulses generated by a Pulse Streamer was utilized, and time-gated measurements were controlled by a Time Tagger at a power of ~2.3 mW. Fluorescence images were captured using the open-source Qudi software, scanning regions of 5 µm x 5 µm with a spatial resolution of 0.1 µm at a power of ~2.3 mW (measured in CW mode). Microwaves generated by a microwave/RF signal generator (Berkeley Nucleonics, Model 835) were delivered to the sample via a printed circuit board (PCB), with the diamond mounted on the microwave resonator using carbon tape. During the ODMR experiments, the microwave frequency was swept from 2.83 GHz to 2.92.

**NV center depths:**

We estimated NV center depth based on nitrogen ion implantation energy using stopping and range of ions in matter (SRIM) simulations. Diamond samples were ion implanted at 4 keV (bare diamond) and 6 keV (diamond with Si antennas). As shown in Fig. S1(a), 4 keV implantation yields an NV center penetration depth centered at ~7 nm, while 6 keV results in a depth centered at ~9.4 nm. For the unpatterned diamond, these depths in this range should not have a big impact on how much light emerges from the diamond (Fig. S1(b)). We chose the 10 nm case as the representative baseline for unpatterned diamond. Fig. S1(c) shows the Purcell-enhanced extraction enhancement, $\mathcal{E}(\lambda) \times F(\lambda)$, of Si antennas with a diameter of 500 nm and a height of 220 nm as a function of NV center depth, indicating no significant difference in extraction enhancement for NV depths between 5 nm and 10 nm.

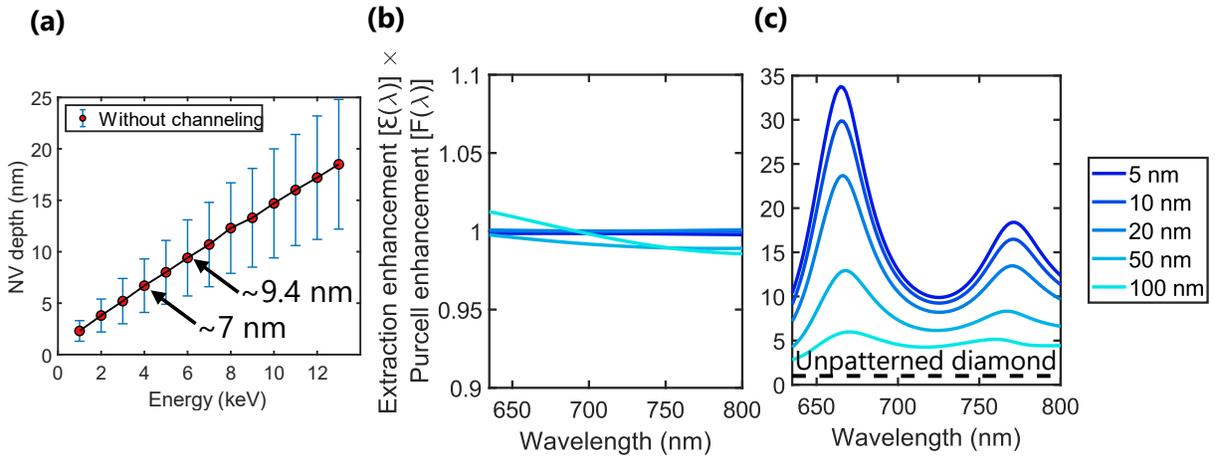

**Figure S1. (a)** Stopping and range of ions in matter (SRIM) simulation, performed assuming no channeling, of nitrogen ions implanted into diamond. **(b)** Purcell-enhanced extraction "enhancement", $\mathcal{E}(\lambda) \times F(\lambda)$, for unpatterned diamond as a function of NV depths, normalized to the case of 10 nm, showing essentially no change. **(c)** $\mathcal{E}(\lambda) \times F(\lambda)$ of Si antennas as function of NV depths, normalized to the unpatterned case with an NV depth of 10 nm.

**Optically detected magnetic resonance (ODMR):**

The ODMR spectra of the NV centers shown in the main text (Fig. 3(a, e)) were measured by sweeping the microwave source over the NV center splitting between the $m_s = 0$ and $m_s = \pm 1$ ground-state levels (Fig. 1(a)) without an external magnetic field. Each spectrum shows a characteristic dip at 2.87 GHz. While the UVO treatment significantly reduced background fluorescence, the ODMR contrast for NVs under the Si pillar remained lower than for NVs in bare diamond (Fig. S2 (b)). This reduced contrast is not directly linked to background fluorescence but is influenced by experimental variations, particularly NV-antenna misalignment and fabrication inconsistencies. These factors affect the coupling between NV centers and the applied microwave field, requiring adjustments in microwave power for each sample. Additional measurements (e.g., Fig. S9(e)) demonstrate that under optimized conditions, the ODMR contrast can approach levels seen in bare diamond.

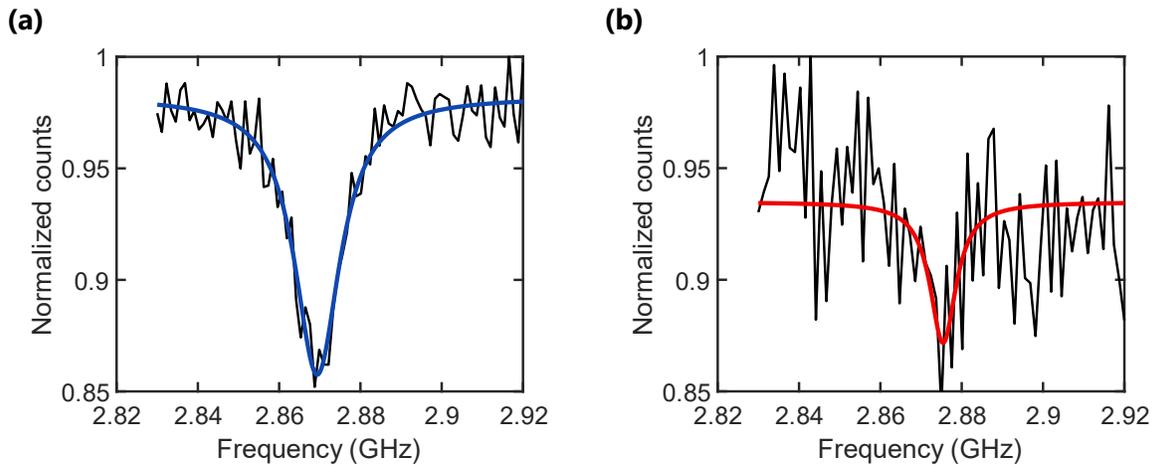

**Figure S2.** Optically detected magnetic resonance (ODMR), showing a dip at 2.87 GHz, from **(a)** bare diamond (Fig. 3(a)) and **(b)** Si antennas on diamond (Fig. 3(e)).

**Reduction of background due to the Si membrane:**

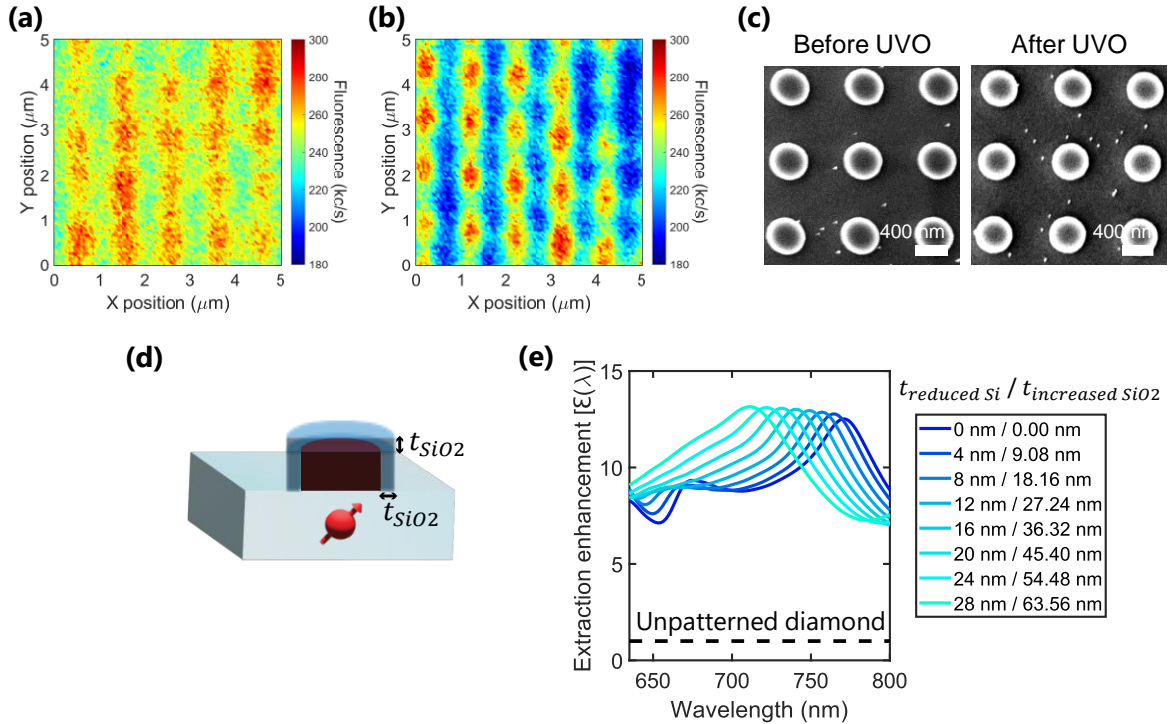

**Figure S3. (a)** Confocal microscopy scan of a 5-μm × 5-μm region of an optical-grade diamond before treatment. In this sample, the NV density is high, and there are multiple NVs beneath every pillar as well as in regions with no pillar. **(b)** Scan of the same region after UV/Ozone (UVO) treatment. **(c)** SEM images of the Si pillars before and after UVO treatment, showing an increase in the overall size of the pillars from 510 nm to 535 nm due to surface oxidation. **(d)** Schematic illustrating the reduction in the diameter and height of the Si region and the growth of the SiO₂ shell due to surface oxidation. The overall size of the antennas increases due to the addition of the SiO₂ shell. **(e)** Simulated extraction enhancement [$\mathcal{E}(\lambda)$] of the Si antenna as a function of decreasing pillar size and increasing SiO₂ shell thickness.

In the initial confocal scans, significant background fluorescence was observed (Fig. S3(a)), which we attribute primarily to fluorescence from the Si membrane. To suppress this background signal, we applied UV/Ozone (UVO) treatment at 200 °C with a continuous flow of 0.5 L/min of O₂ for 35 minutes.[1] After treatment, the background fluorescence was significantly reduced, as shown in Fig. S3(b). However, the overall size of the pillars increased due to the conversion of Si to SiO₂ during the UVO treatment (Fig. S3(c)). This process reduced the size of the Si pillar while forming a SiO₂ shell, with the volume expansion factor (VEF) of 2.27 resulting in, for example, a 4 nm reduction in Si thickness leading to a SiO₂ shell thickness of approximately 9.08 nm. To better understand this process, we simulated the impact of Si reduction and SiO$_2$ shell formation on the extraction enhancement of the NV center's emission. A schematic of the process is shown in Fig. S3(d), where the Si antenna shrinks in diameter and height while the SiO$_2$ shell thickness grows. The simulated results in Fig. S3(e) show that the extraction enhancement decreases slightly and shifts to shorter wavelengths as the Si pillar dimensions shrink and the SiO$_2$ shell thickness grows. For the electronic-grade diamond with isolated NVs used in the main text, we applied UVO treatment at 200 °C for 10 minutes, instead of 35 minutes like in Fig. S3(b,c). This duration effectively reduced the overall background fluorescence while minimizing any changes to the pillar dimensions.

**Elongated shape of the Si pillar in confocal measurement:**

In the confocal scans such as the ones in Fig. 3(a, e) and S4(a), the pillars appear elongated along the y-direction. This elongation persisted in the y-direction even after rotating the sample. Adjusting the alignment of the confocal microscopy mitigated this issue (Fig. S4(b)). However, the alignment is highly sensitive to vibrations and external environmental factors, and unfortunately our laboratory had a lot of vibration at the time from construction outside. Therefore, some of the scans in the main text and the supplementary still have a bit of asymmetry.

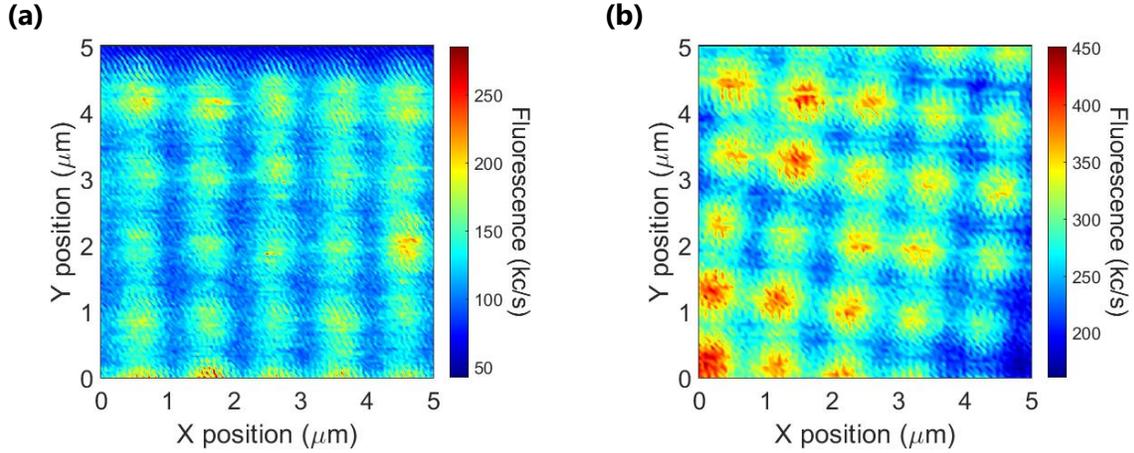

**Figure S4. (a)** Confocal microscopy scan of a 5-µm × 5-µm region of an optical-grade diamond with high NV density before alignment adjustments. The pillars enhance emission from multiple NVs. **(b)** Confocal microscopy scan of a 5-µm by 5-µm region of the same sample after alignment adjustments.

**NV center and Si antenna alignment:**

Prior to fabrication, we performed confocal microscopy on a diamond sample ion-implanted at 4 keV, identifying 14 potential single NV centers within a 10 µm × 10 µm area. Using this information, we simulated a square array of Si pillars—each 500 nm in diameter with a 1 µm period—randomly positioned on the NV centers. A 'hit' was defined as an instance where the distance between the center of a Si pillar and an NV center is less than 150 nm. After 1000 simulation iterations, we found an ~8% probability of a Si pillar aligning with an NV center within a 10 µm × 10 µm area. This suggests that across 13 scans of such 10 µm × 10 µm regions with similar densities as the confocal scan in Fig. S5(a), approximately one Si pillar is likely to be properly aligned with an NV center.

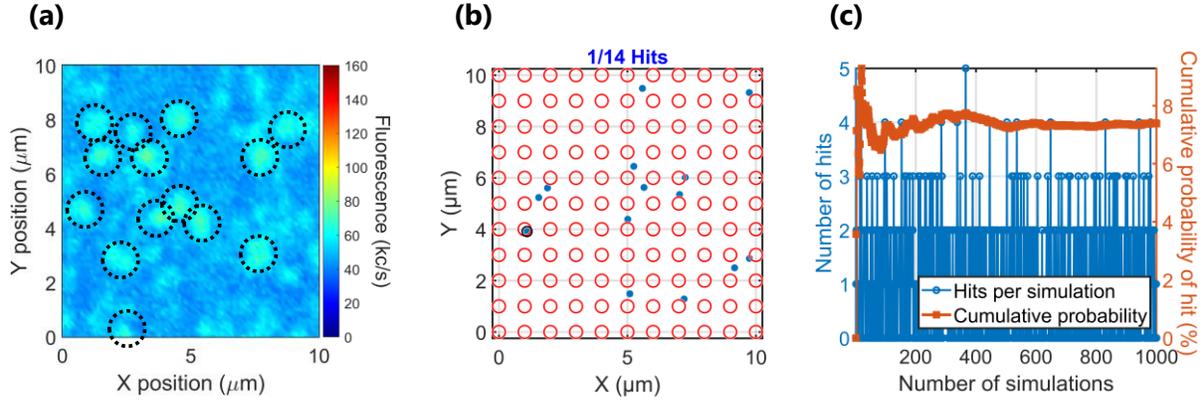

**Figure S5. (a)** Confocal scan of the electronic-grade diamond ion-implanted with 4 keV before Si membrane transfer, with black circles representing potential single NVs. **(b)** Simulation of a square array of Si pillars (500 nm diameter, 1 μm period) showing the probability of a Si pillar being positioned on top of NVs (blue dots). A 'hit' is defined as a scenario where the distance between the center of a Si pillar and an NV is less than 150 nm. **(c)** Simulated cumulative probability of achieving a hit, based on 1000 iterations.

Since we did not intentionally align NV centers and Si antennas during e-beam lithography exposure, the antennas and NVs were almost always at least partially misaligned (Fig. S6(a)). To simulate the effects of potential misalignment of the NV center, we performed FDTD simulations. Fig. S6(b) and Fig. S6(c) show the NV center position along the x-axis, ranging from 0 nm (aligned) to 250 nm (at the edge of the pillar), with the NV center depth fixed at 10 nm. The extraction efficiency (Fig. S6(b)) decreases with increasing misalignment, and shape of the Purcell enhancement spectrum (Fig. S6(c)) changes, though the spectrum-averaged value does not change too much.

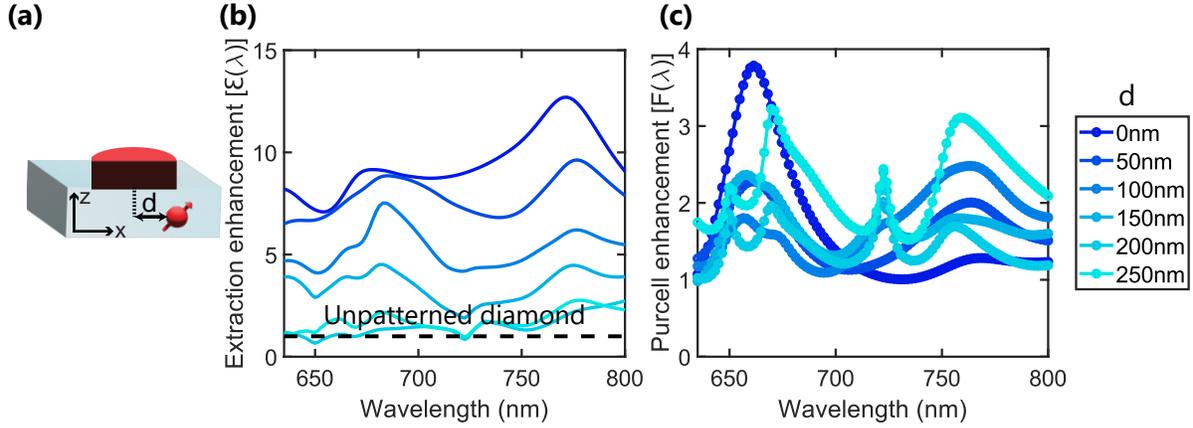

**Figure S6. (a)** Schematic of the misalignment of the Si antenna and NV center at a depth of 10 nm. **(b)** Simulated extraction enhancement [$\mathcal{E}(\lambda)$] of the Si antenna as a function of NV center misalignment, ranging from 0 nm to 250 nm. **(c)** Purcell enhancement [$F(\lambda)$] of the Si antenna as a function of the same misalignment.

Fig. S7(b-d) shows the laser power perceived by NV centers located 5 nm, 10 nm, and 20 nm below the surface. The intensity under the antenna change significantly as a function position, and is not rotationally symmetric due to the linear incident polarization. This nontrivial intensity distribution, combined with the fact that in our experiment we did not intentionally control the incident polarization, means that we cannot evaluate the exact incident intensity at the NV center from one sample to another, which means that our values of $P_{sat}$ should not be compared from one measurement to the next. An improvement to future experiments would be to deliberately control the incident polarization, and acquire scans for different incident polarizations.

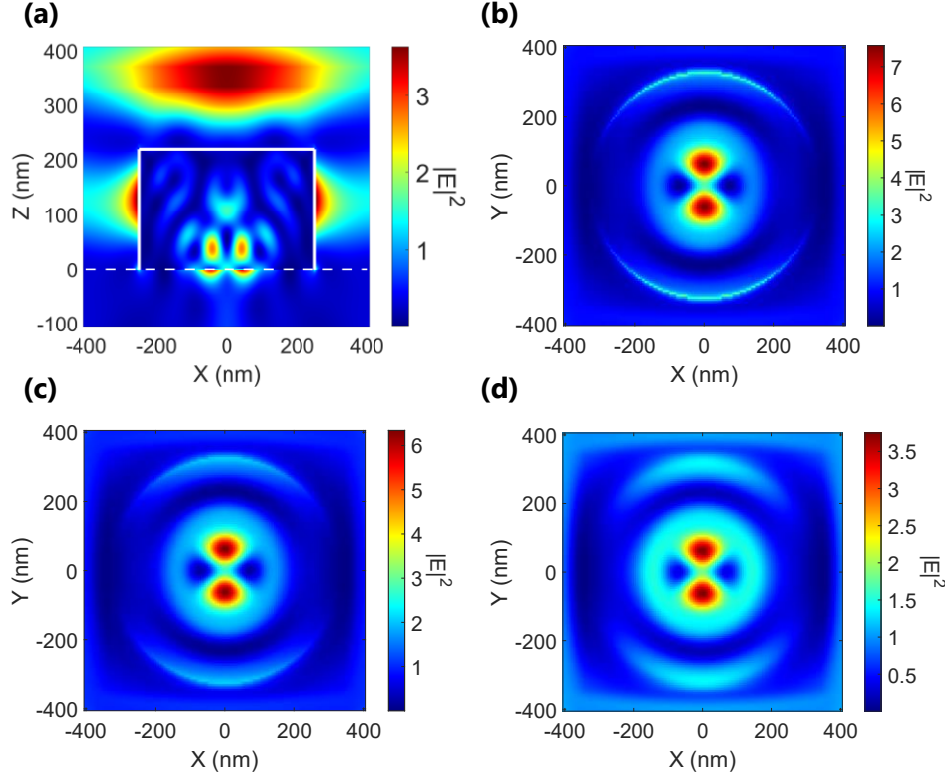

**Figure S7. (a)** Snapshot of the electric field intensity, $|E|^2$, when an x-polarized 515 nm laser passes is incident onto the through the Si antenna and diamond at normal incidence to the diamond surface. **(b-d)** Simulated laser intensity perceived by NV centers located at different depths below the surface when illuminated by a 515 nm laser, as a function of x- and y- coordinates. **(b)** 5 nm below the surface. **(c)** 10 nm below the surface. **(d)** 20 nm below the surface.

**Measurements of additional samples not shown in the main text:**

In this section, we show complete measurements for four additional antenna-enhanced NVs (Fig. S8-11). The enhancement in these measurements is not as high as the sample in the main text (total enhancement of ~4 vs. ~9 in the main text), likely due to worse alignment between the antenna and the NV. For the sample in Fig. S8, we also performed $T_2$ measurements.

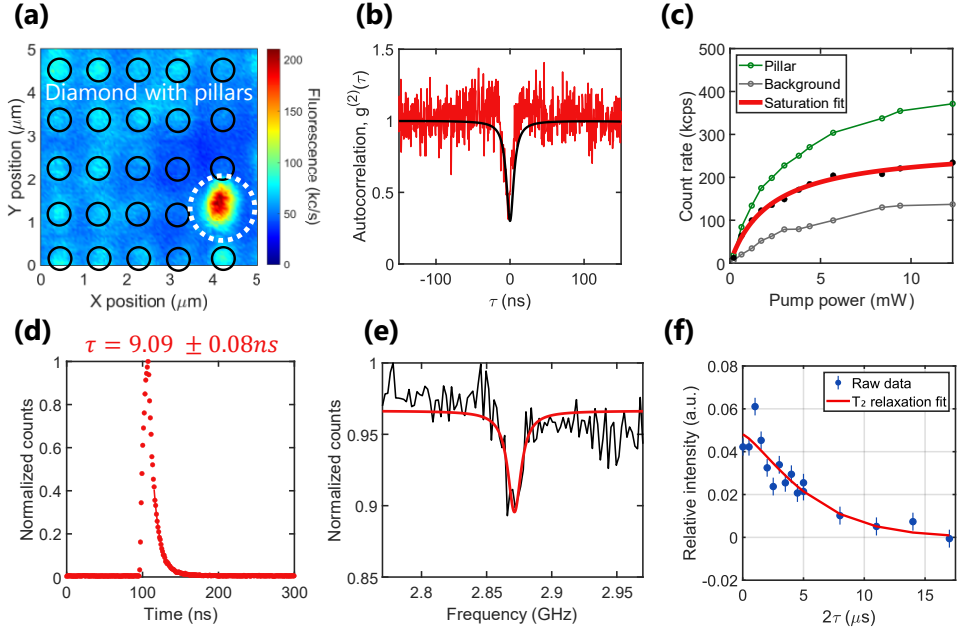

**Figure S8.** Si antenna showing a 4.07x enhancement. **(a)** Confocal microscopy scans of a 5-μm by 5-μm region, showing circled regions (white dashed circles) used for measurement and black circles representing other nearby pillars. **(b)** Autocorrelation function $g^2(0) < 0.5$ for a single NV underneath a Si antenna. **(c)** Saturation curve of the total count rates as a function of pump power. The background is subtracted from the total signal, and the isolated NV emission (black dots) is fitted to a saturation model (red line). **(d)** Normalized fluorescence decay for a single NV center with a Si antenna, fitted to a one-exponential model. The fit yields a time constant of 9.09 ± 0.08 ns. **(e)** ODMR showing a dip at 2.87 GHz from the Si antenna on diamond. **(f)** $T_2$ relaxation measurement for the Si antenna, with raw data (blue points) fitted to a $T_2$ relaxation model (red line). This $T_2$ is consistent with the reported values for near-surface NV centers of similar depth (~7 nm).[2]

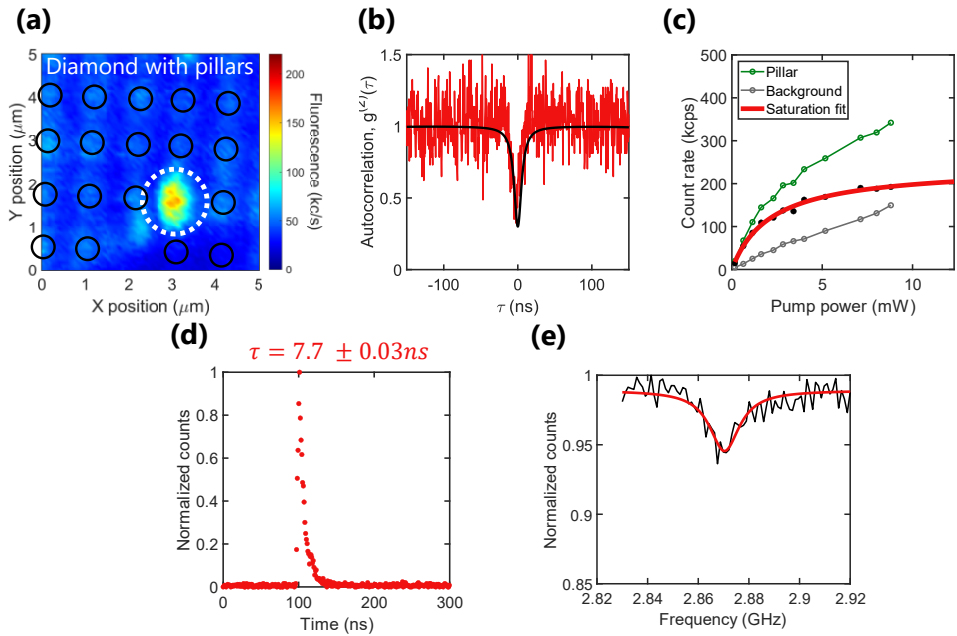

**Figure S9.** Si antenna showing a 3.59x enhancement. **(a)** Confocal microscopy scans of a 5-µm by 5-µm region, showing circled regions (white dashed circles) used for measurement and black circles representing other nearby pillars. **(b)** Autocorrelation function $g^2(0) < 0.5$ for a single NV underneath a Si antenna. **(c)** Saturation curve of the total count rates as a function of pump power. The background is subtracted from the total signal, and the isolated NV emission (black dots) is fitted to a saturation model (red line). **(d)** Normalized fluorescence decay for a single NV center with a Si antenna, fitted to a one-exponential model. The fit yields a time constant of 7.7 ± 0.03 ns. **(e)** ODMR showing a dip at 2.87 GHz from the Si antenna on diamond.

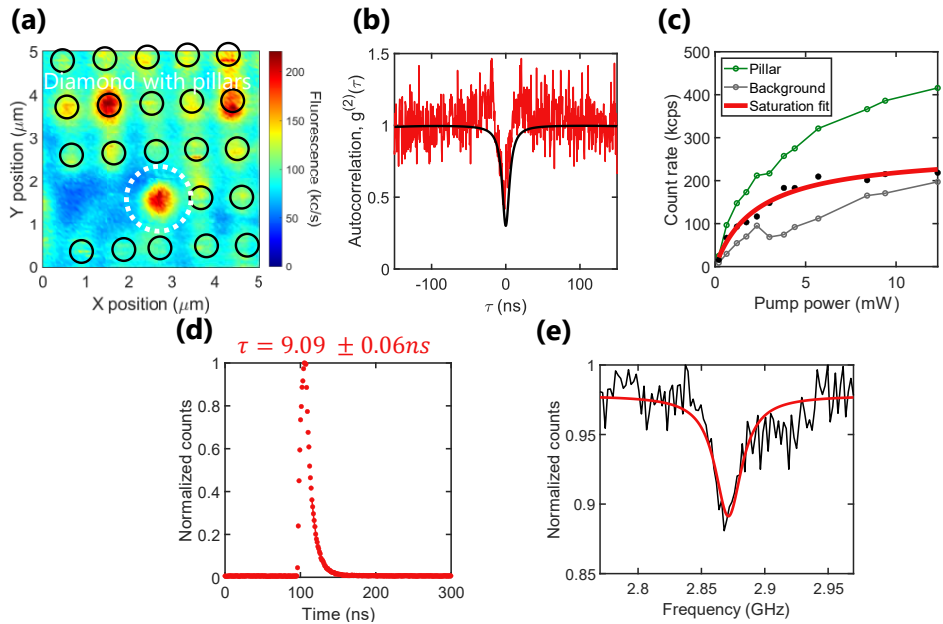

**Figure S10.** Si antenna showing a 4.02x enhancement. **(a)** Confocal microscopy scans of a 5-µm by 5-µm region, showing circled regions (white dashed circles) used for measurement and black circles representing other nearby pillars. **(b)** Autocorrelation function $g^2(\tau)$ for a single NV underneath a Si antenna. **(c)** Saturation curve of the total count rates as a function of pump power. The background is subtracted from the total signal, and the isolated NV emission (black dots) is fitted to a saturation model (red line). **(d)** Normalized fluorescence decay for a single NV center with a Si antenna, fitted to a one-exponential model. The fit yields a time constant of 9.09 ± 0.06 ns. **(e)** ODMR showing a dip at 2.87 GHz from the Si antenna on diamond.

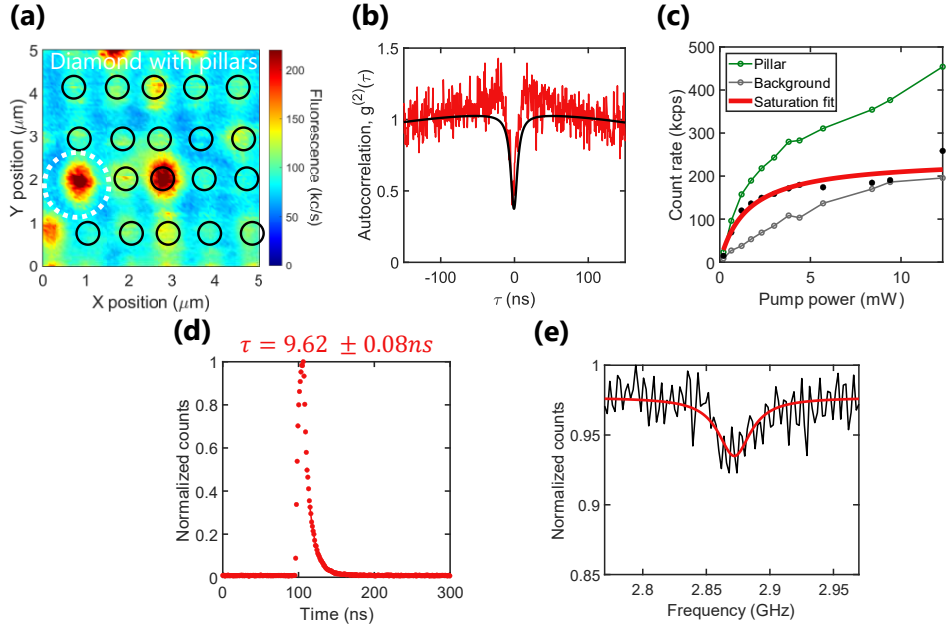

**Figure S11.** Si antenna showing a 3.6x enhancement. **(a)** Confocal microscopy scans of a 5-μm by 5-μm region, showing circled regions (white dashed circles) used for measurement and black circles representing other nearby pillars. **(b)** Autocorrelation function $g^2(\tau)$ for a single NV underneath a Si antenna. **(c)** Saturation curve of the total count rates as a function of pump power. The background is subtracted from the total signal, and the isolated NV emission (black dots) is fitted to a saturation model (red line). **(d)** Normalized fluorescence decay for a single NV center with a Si antenna, fitted to a one-exponential model. The fit yields a time constant of 9.62 ± 0.08 ns. **(e)** ODMR showing a dip at 2.87 GHz from the Si antenna on diamond.

**Possible fabrication deviations from the design, besides lateral misalignment:**

To prevent diamond surface damage from $O_2$ plasma etching[3] and minimize unwanted contamination, especially for near-surface NV centers (~10 nm), we intentionally left a thin layer of Si on top of the diamond during the etching process. Based on the calibrated etching rate, we estimate that a Si membrane thickness of 20–30 nm remains on the diamond. Simulations show that the extraction enhancement shows no significant difference for Si layers up to ~40 nm (Fig. S12).

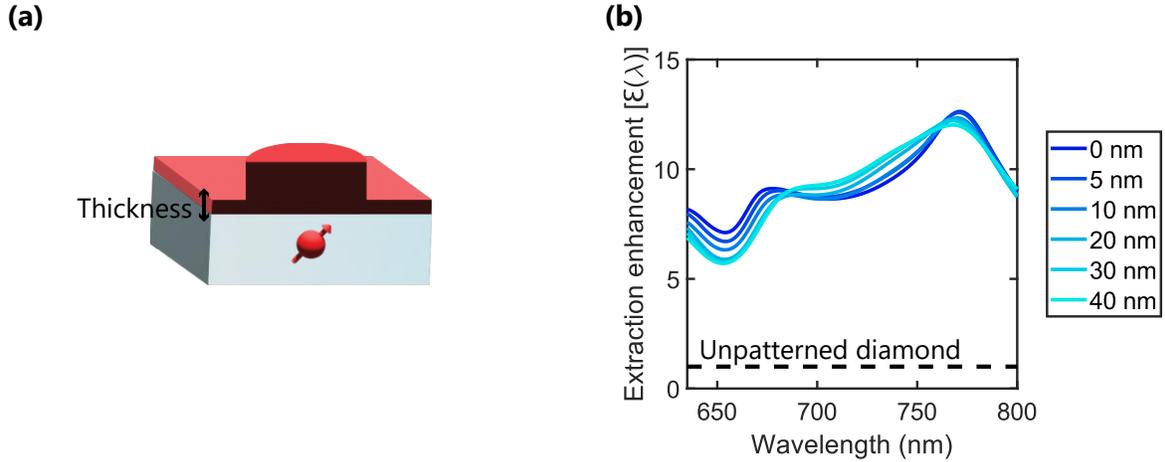

**Figure S12. (a)** A schematic showing intentional under-etching the silicon (Si) membrane on top of the diamond. **(b)** Simulated extraction enhancement [$\mathcal{E}(\lambda)$] due to the Si antenna as a function of the remaining thickness of the Si membrane atop the diamond.

To define the circular shape of Si antenna, we used an $SiO_2$ hard mask as the etch mask (Fig. S13(a)). After etching, some $SiO_2$ remained on top of Si antennas. We chose to not remove this layer, because it is not expected to affect the optical properties due to its low refractive index (n ~ 1.45) and small thickness. This is confirmed in the simulations of extraction enhancement as a function of the $SiO_2$ layer on top (Fig. S13(b)).

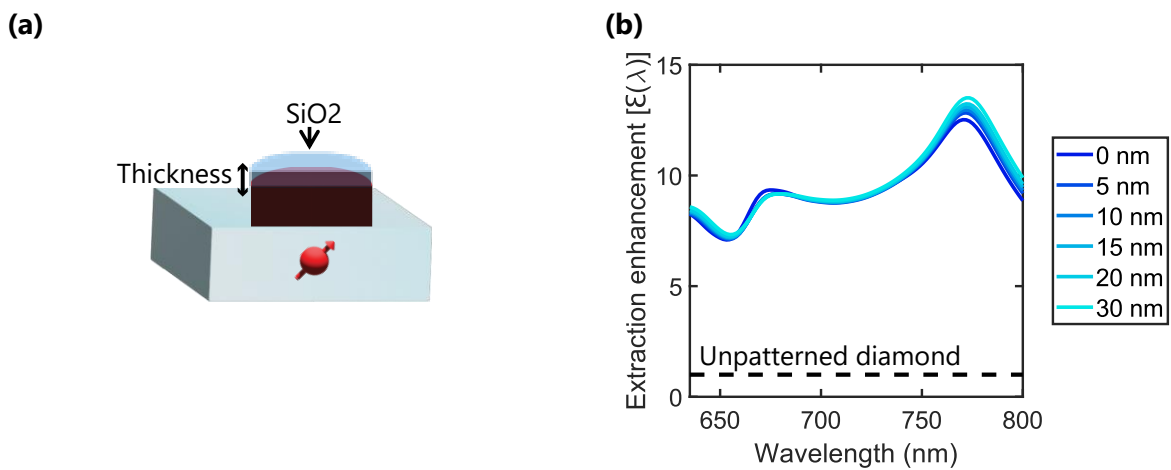

**Figure S13. (a)** A schematic of the $SiO_2$ hard mask on top of the Si antenna. **(b)** Simulated extraction enhancement [$\mathcal{E}(\lambda)$] of the Si antenna as a function of the remaining thickness of the $SiO_2$ hard mask.

In the SEM image (Fig. 2), slightly tilted sidewalls of Si antennas are observed, as illustrated in Fig. S14(a). The average reduction in the top radius of the Si antenna compared to the bottom, denoted here as $x$, is ~17 nm, as measured from the SEM. Fig. S14(b) shows that the tilted sidewalls shift the resonance peak of the Si antenna but have only a small impact on its overall performance.

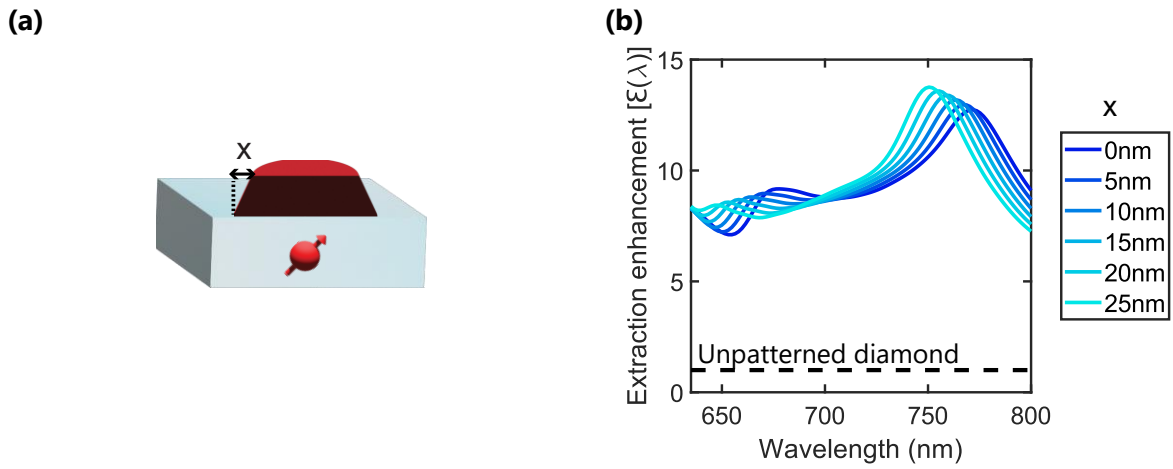

**Figure S14. (a)** A schematic of the Si antenna with a tilted sidewall. **(b)** Simulated extraction enhancement, $\mathcal{E}(\lambda)$, of the Si antenna as a function of the reduction in the pillar's top radius.

**Si antenna with an inner hole:**

We note that a Si antenna positioned directly on top of an NV may be detrimental for certain sensing applications where an analyte needs to be close to the NV, for example in chemical sensing.[4] To confirm that our approach remains valid when direct access to the diamond surface is needed, we simulated a 220 nm tall Si pillar with an internal hole, sweeping over the inner diameter (d) and outer diameter (D) (Fig. S15(a)). We observe that an outer diameter of 500 nm and an inner diameter of 50 nm can achieve nearly a 10-fold increase in light extraction from an NV center while providing direct access to the diamond surface. This design enables analytes or other components to interact directly with the diamond surface without compromising light extraction enhancement.

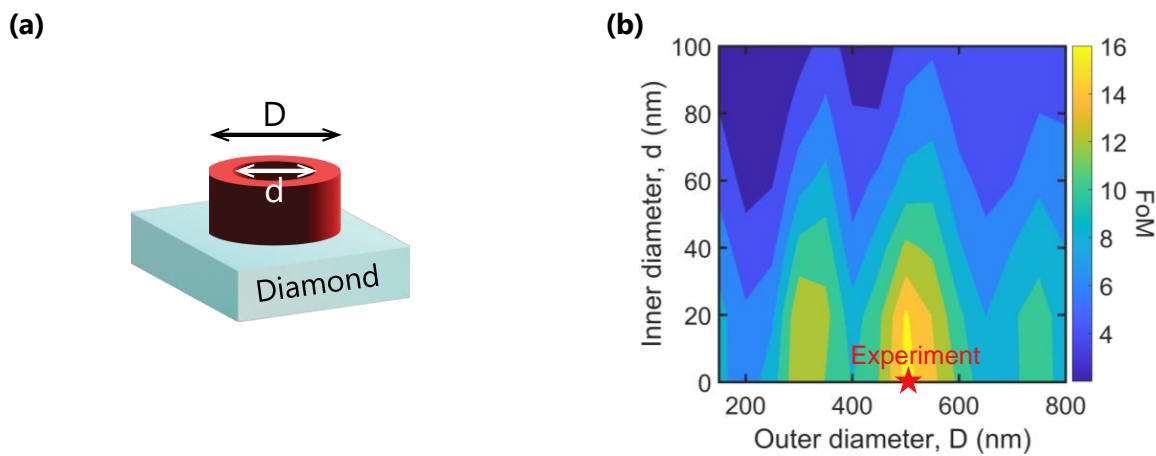

**Figure S15. (a)** Schematic of the Si antenna with an inner hole, defined by its outer diameter $D$ and inner diameter $d$. **(b)** Color map of the FoM with respect to the outer diameter ($D$) and inner diameter ($d$) of the Si antenna. The simulation result that corresponds to the experiments in Figs. 2-4 is marked with a red star.

**Supplementary references**